\documentclass[aps,prb,twocolumn,showpacs,superscriptaddress]{revtex4}

\usepackage{graphicx}
\usepackage{bm}
\usepackage{amsmath}
\usepackage{dcolumn}%

\newcommand{\dxy}{\ensuremath{d_{xy}}}
\newcommand{\dxz}{\ensuremath{d_{xz}}}
\newcommand{\dyz}{\ensuremath{d_{yz}}}
\newcommand{\dzz}{\ensuremath{d_{3z^2-1}}}
\newcommand{\dxxyy}{\ensuremath{d_{x^2-y^2}}}

\newcommand{\ef}{\ensuremath{\varepsilon_{F}}}

\newcolumntype{/}{D{/}{/}{2,2}}  
\newcolumntype{.}{D{.}{.}{0}}  

\begin{document}

\title{Comparative study of the electronic structure, phonon spectra
  and electron-phonon interaction of ZrB$_2$ and TiB$_2$}

\author{S.M. Sichkar}
\affiliation{Institute of Metal Physics, 36 Vernadsky Street, 03142
Kiev, Ukraine}

\author{V.N. Antonov}

\affiliation{Institute of Metal Physics, 36 Vernadsky Street, 03142
Kiev, Ukraine}

\affiliation{Ames Laboratory USDOE, Ames, IA 50011}

\author{V.P. Antropov}

\affiliation{Ames Laboratory USDOE, Ames, IA 50011}

\date{\today}

\begin{abstract}

The electronic structure, optical and x-ray absorption spectra, angle
dependence of the cyclotron masses and extremal cross sections of the Fermi
surface, phonon spectra, electron-phonon Eliashberg and transport spectral
functions, temperature dependence of electrical resistivity of the MB$_2$
(M=Ti and Zr) diborides were investigated from first principles using the full
potential linear muffin-tin orbital method. The calculations of the dynamic
matrix were carried out within the framework of the linear response theory. A
good agreement with experimental data of optical and x-ray absorption spectra,
phonon spectra, electron-phonon spectral functions, electrical resistivity,
cyclotron masses and extremal cross sections of the Fermi surface was
achieved.

\end{abstract}


\maketitle

\section{\label{sec:introd}Introduction}

Ceramics based on transition metal borides, nitrides, and carbides have
extremely high melting points ($>$2500 $^{\circ}$C) and are referred to as
ultra-high temperature ceramics. \cite{UYH97,FHT+07} Among them, diborides
such as ZrB$_2$ and HfB$_2$ have a unique combination of mechanical and
physical properties: high melting points ($>$3000 $^{\circ}$C); high thermal
and electrical conductivity; chemical inertness against molten metals; great
thermal shock resistance.  \cite{UYH97,FHT+07,Mroz94} Thus, although carbides
typically have the highest melting points ($>$3500 $^{\circ}$C), the diborides
ZrB$_2$ and HfB$_2$ are more attractive candidates for high-temperature
thermomechanical structural applications at temperatures $\geq$3000
$^{\circ}$C. \cite{UYH97,FHT+07} Potential applications include thermal
protective structures for leading-edge parts on hypersonic re-entry space
vehicles, \cite{UYH97,Brown97} propulsion systems, \cite{UYH97,Brown97}
furnace elements, \cite{Kuw02} refractory crucibles, \cite{Kuw02} and
plasma-arc electrodes.  \cite{Kuw02,NEB+99} In particular, ZrB$_2$ has the
lowest theoretical density among the ultra-high temperature ceramics, which
makes it an attractive material for aerospace applications.
\cite{UYH97,FHT+07,Brown97} Titanium diboride is also potentially useful
because it has many interesting physical properties, such as low density and
unusual strength. \cite{LMM+96} TiB$_2$ is widely accepted for applications
including microelectronics, diffusion barriers, wear- and erosion-resistant
coatings for cutting tools and other mechanical components. In these
applications, the material's high hardness, high melting point, good
electrical conductivity, and acid and radiation stability is
exploited. \cite{Munro00}

The discovery of superconductivity in MgB$_2$ at 39 K by Akimitsu
\cite{NNM+01} has lead to booming activity in the physics community and
activated a search for superconductivity in other diborides.  Natural
candidates for this search are AB$_2$-type light metal diborides (A = Li, Be,
Al). However, up to now superconductivity has not been reported in the
majority of these compounds. \cite{GSZ+01} Only very recently has
superconductivity below 1 K ($T_c$ = 0.72 K) been reported in
BeB$_{2.75}$. \cite{cm:YAC+01} According to Ref.  \onlinecite{BuYa01} no
superconducting transition down to 0.42 K has been observed in powders of
diborides of transition metals (A = Ti, Zr, Hf, V, Ta, Cr, Mo, U). Only
NbB$_2$ is expected to superconduct with a rather low transition temperature
($<$ 1 K), and contradictory reports about superconductivity up to $T_c$=9.5 K
in TaB$_2$ can be found in Ref. \onlinecite{BuYa01}. Finally, the reported
$T_c$=7 K in ZrB$_2$ \cite{GSZ+01} encourages further studies of these
diborides.

Presently, a number of experimental studies exist dealing with the physical
properties of ZrB$_2$ and TiB$_2$ such as electric transport properties,
\cite{LMM+96,GKS+04,FCP+06,GaSu06,ZHF08,ZPM11} the de Haas-van Alphen (dHvA)
measurements of the Fermi surface, \cite{TIB+78,IsTa86,TaIs+80,PSD+07} optical
\cite{Linton74,RDA+08,RAC+08} and electron-energy-loss spectra,
\cite{LBD99,LHB00} x-ray absorption and photoemission spectra,
\cite{IHN77,TEK+03} magnetic susceptibility \cite{GFL+09,FGP+09} and NMR
measurements, \cite{LuLa05} the phonon density of states, \cite{HRS+03} and
electron-phonon interaction. \cite{NKY+02,HRS+03} First-principles
calculations of the electronic structure of diborides have been widely
presented.
\cite{IHN77,JHL80,BCM86,AnDo90,LBD99,LHB00,VRR+01,RAP+02,MCT02,ShIv02,Pad03,Singh04,DRO+04,MII04,MFA06,HDS+07,ZLH+08,FGP+09,DCC09,ZLL+09,DCC10,FTH+10}

Despite a lot of publications, there are still many open questions related to
the electronic structure and physical properties of transition metal
diborides. Ihara {\it et al.} \cite{IHN77} calculated the band structure and
the density of states (DOS) of ZrB$_2$ by using an augmented plane wave
method. They pointed out that the band structure of ZrB$_2$ is determined by
the $sp^2$ hybridization, $p_z$ state of B and the 4$d$ and 5$s$ states of
Zr. Similarly Johnson \cite{JHL80} calculated the band structure of ZrB$_2$
using the Korringa-Kohn-Rostoker method in the spherical muffin-tin
approximation. However, they concluded that the B 2$s$ states are localized
and do not hybridize with B 2$p$. Pablo {\it et al.}  \cite{MCT02} compared
the electronic structure of iso-structural alkaline-earth diborides using a
full-potential linearized augmented plane wave (FLAPW) method and found that
Zr-B bonds have covalent character, yet still remain highly ionic. Fermi
surfaces and DOS values at the Fermi level reported by Shein {\it et al.}
\cite{ShIv02} and those by Rosner {\it et al.}  \cite{RAP+02} are quite
different. Vajeeston {\it et al.}  \cite{VRR+01} also investigated the
electronic structure of AlB$_2$-type diborides using the tight-bonding linear
muffin-tin orbital (TB-LMTO) method, they claimed that metal-metal and
metal-boron interactions are less significant than the $p-p$ covalent
interaction of boron atoms. Burdett {\it et al.}, \cite{BCM86} on the basis of
orbital overlap, indicated the importance of the interaction of orbitals of
the metal with those of a graphite-like net of boron atoms as well as the
interaction with those of other metals in influencing the properties of these
species. The bonding nature, elastic property and hardness were investigated
by Zhang {\it et al.}  \cite{ZLH+08} for ZrB$_2$ using the plane-wave
pseudopotential method. The stiffness and the thermal expansion coefficient of
ZrB$_2$ were calculated using the density functional theory formalism by
Milman {\it et al.} in Ref. \onlinecite{MWP05}. Kaur {\it et al.}
\cite{KMG+09} studied the cohesive and thermal properties of these compounds
using the rigid ion model. The elastic properties, electronic structure,
electronic charge distribution, and equation of states of titanium diboride
were studied by Milman and Warren, \cite{MiWa01} Perottoni {\it et al.},
\cite{PPJ00} and Camp {\it et al.}  \cite{CaDo05} using the first-principles
methods. Peng {\it et al.}  \cite{PFC07} investigated the thermodynamic
properties of TiB$_2$ using a plane-wave pseudopotential method. Munro
\cite{Munro00} examined the physical, mechanical, and thermal properties of
polycrystalline TiB$_2$ and showed that these properties are significantly
related to the density and grain size of the used specimens. Deligoz {\it et
  al.}  \cite{DCC09,DCC10} investigated the structural and lattice dynamical
properties of TiB$_2$ and ZrB$_2$ together with VB$_2$, ScB$_2$, NbB$_2$, and
MoB$_2$. They specifically presented following properties: lattice parameters;
bond lengths; phonon dispersion curves and corresponding density of states;
some thermodynamic quantities such as internal energy, entropy, heat capacity,
and their temperature-dependent behaviors. Systematic trends in lattice
constants and heats of formation for these compounds were studied by
Oguchi. \cite{Oguchi02} Vajeeston {\it et al.}  \cite{VRR+01} investigated the
electronic structure and ground state properties of these diborides using
TB-LMTO. X-ray absorption and photoemission spectra of ZrB$_2$ and TiB$_2$
were measured experimentally in Refs. \onlinecite{IHN77,TEK+03,CWA+05}. The
optical spectra of ZrB$_2$ was investigated experimentally by several authors,
\cite{Linton74,RDA+08,RAC+08} however, there is neither theoretical
investigation of the x-ray absorption spectra or the optical properties of the
transition metal diborides.

The band structure and Fermi surface parameters were studied by Shein and
Ivanovskii \cite{ShIv02} using the self-consistent full potential linearized
muffin-tin orbital (FP-LMTO) method for ZrB$_2$ and NbB$_2$. Rosner {\it et
  al.}  \cite{RAP+02,DRO+04} provided a comparison of full potential band
calculations of the Fermi surfaces areas and masses of MgB$_2$ and ZrB$_2$
with dHvA data for several symmetry points in the Brillouin zone (BZ). They
found, with one possible exception, that LDA provides a good description for
ZrB$_2$. For MgB$_2$ some disagreement in FS areas can be accounted for by a
shift of $\pi$ (B $p_z$) bands with respect to $\sigma$ (B $sp_xp_y$) bands by
240 meV and by a readjustment of the "Fermi energies" of each of these bands
by $\pm$120 meV. Heid {\it et al.} \cite{HRS+03} measured the phonon density
of states of MB$_2$ with M=Ti, V, Ta, Nb, and Y using inelastic neutron
scattering. Experimental data were compared with {\it ab initio}
density-functional calculations using the mixed basis pseudopotential method.
The results do not exhibit indications of strong electron-phonon interaction
in the diborides considered.  Singh \cite{Singh04} studied electron-phonon
interaction in ZrB$_2$ and TaB$_2$ using a FP-LMTO method. The results for
phonon density of states and Eliashberg function show electron-phonon coupling
in ZrB$_2$ to be much weaker than in TaB$_2$. The average electron-phonon
coupling constant $\lambda$ is found to be 0.15 for ZrB$_2$ and 0.73 for
TaB$_2$. Solutions of the isotropic Eliashberg gap equation indicate no
superconductivity for ZrB$_2$.

The aim of this work is a complex comparative investigation of the electronic
structure, optical and x-ray absorption spectra, angle dependence of the
cyclotron masses and extremal cross sections of the Fermi surface, phonon
spectra, electron-phonon interaction and electrical resistivity of the
diborides TiB$_2$ and ZrB$_2$. The paper is organized as follows. Section
\ref{sec:details} presents the details of the calculations. Section
\ref{sec:results} is devoted to the electronic structure as well as optical
and x-ray absorption spectra, angle dependence of the cyclotron masses and
extremal cross sections of the Fermi surface, phonon spectra, electron-phonon
interaction and electrical resistivity using the FP-LMTO band structure
method. The results are compared with available experimental data. Finally,
the results are summarized in Sec.~\ref{sec:summ}.

\begin{figure}[tbp!]
\begin{center}
\includegraphics[width=0.99\columnwidth]{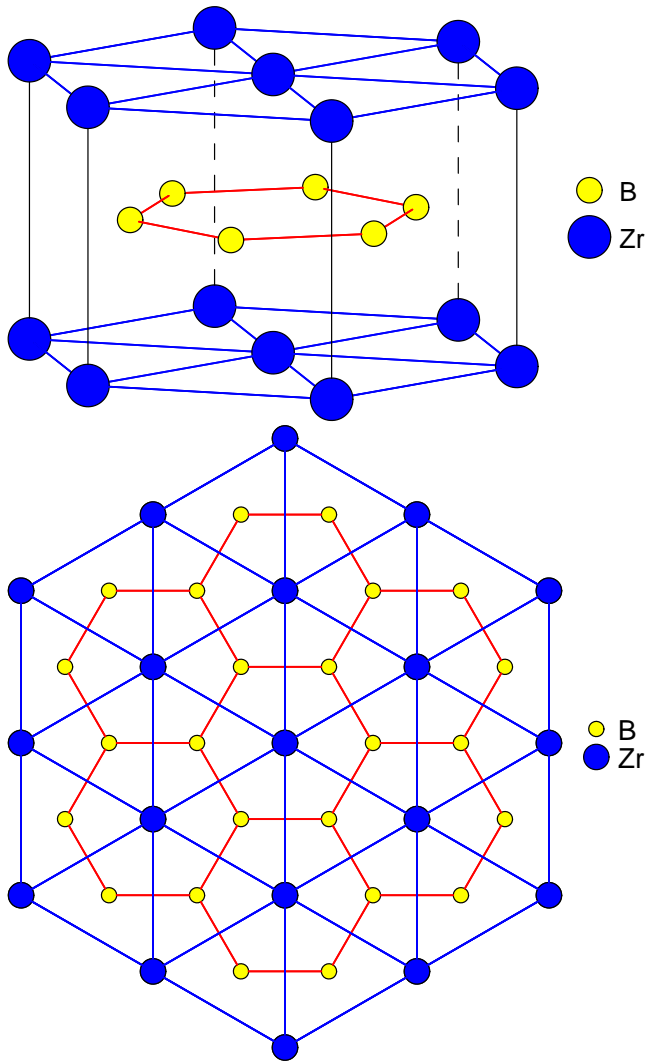}
\end{center}
\caption{\label{struc}(Color online) Schematic representation of the
  ZrB$_2$ structure (upper panel) and top view of Zr (large blue
  circles) and B (small yellow circles) planes in ZrB$_2$ (lower
  panel). }
\end{figure}

\section{\label{sec:details}Computational details}

Most known transition-metal (M) diborides MB$_2$ are formed by group III-VI
transition elements (Sc, Ti, Zr, Hf, V, Nb, and others) and have a layered
hexagonal C32 structure of the AlB$_2$-type with the space group symmetry
$P6/mmm$ (number 191). It is simply a hexagonal lattice in which
closely-packed transition metal layers are present alternative with
graphite-like B layers (Fig.~\ref{struc}). These diborides cannot be exactly
layered compounds because the inter-layer interaction is strong even though
the M layers alternate with the B layers in their crystal structure. The boron
atoms lie on the corners of hexagons with the three nearest neighbor boron
atoms in each plane. The M atoms lie directly in the centers of each boron
hexagon, but midway between adjacent boron layers. Each transition metal atom
has twelve nearest neighbor B atoms and eight nearest neighbor transition
metal atoms (six are on the metal plane and two out of the metal plane). There
is one formula unit per primitive cell and the crystal has simple hexagonal
symmetry ($D6h$). By choosing appropriate primitive lattice vectors, the atoms
are positioned at M (0,0,0), B ($\frac{1}{3}, \frac{1}{6}, \frac{1}{2}$), and
B ($\frac{2}{3}, \frac{1}{3}, \frac{1}{2}$) in the unit cell. The distance
between M-M is equal to $c$. This structure is quite close packed, and can be
coped with efficiently and accurately by the atomic sphere approximation
method. However, for precise calculation of the phonon spectra and
electron-phonon interaction, a full potential approximation should be used.

For a crystal where both the fourfold axis and the magnetization $\rm{ \bf M}$
are perpendicular to the sample surface, and the $z$-axis is chosen to be
parallel to them, the dielectric tensor is composed of the diagonal
$\varepsilon_{xx}$ and $\varepsilon_{zz}$, and the off-diagonal
$\varepsilon_{xy}$ components in the form \cite{book:AHY04}

\begin{equation}
\mbox{\boldmath$\varepsilon$}=\left(
\begin{array}{ccc}
\varepsilon_{xx} & \mbox{$\varepsilon_{xy}$} & 0 \\
\mbox{$-\varepsilon_{xy}$} & \mbox{$\varepsilon_{xx}$} & 0 \\
0 & 0 & \mbox{$\varepsilon_{zz}$}
\end{array}
\right) .
\label{defeps_pol}
\end{equation}

The various elements $\hat{\varepsilon}_{\alpha\beta}$ are composed of
real and imaginary parts as follows:
$\hat{\varepsilon}_{\alpha\beta}=\varepsilon_{\alpha\beta}^{(1)}+ i
\varepsilon_{\alpha\beta}^{(2)}$, where $\alpha ,\beta \equiv x,y,z$;
$\varepsilon_{xx}=(n + i k)^{2}$; $n$ and $k$ are the refractive index
and extinction coefficient, respectively. The optical conductivity
tensor $\hat{\sigma}_{\alpha\beta}=\sigma_{\alpha\beta}^{(1)}+ i
\sigma_{\alpha\beta}^{(2)}$ is related to the dielectric tensor
$\varepsilon _{\alpha \beta }$ through the equation
\begin{equation}
 \hat{\varepsilon}_{\alpha
  \beta} (\omega ) = \delta_{\alpha \beta} +\frac{4 \pi
  i}{\omega}\hat{\sigma}_{\alpha\beta} (\omega ).
\label{defsigeps}
\end{equation}

The optical conductivity of ZrB$_2$ and TiB$_2$ has been computed from
the energy band structure by means of the Kubo-Greenwood \cite{Kub57}
linear-response expression: \cite{WC74}

\begin{eqnarray}
\sigma _{\alpha \beta }(\omega ) &=&\frac{-ie^{2}}{m^{2}\hbar
 V_{uc}}\times 
\nonumber \\
&&\sum_{{\bf k}}\sum_{nn^{\prime }}\frac{f(\epsilon
 _{n{\bf k}})-f(\epsilon
_{n^{\prime }{\bf k}})}{\omega _{nn^{\prime }}({\bf k})}\frac{\Pi
_{n^{\prime }n}^{\alpha }({\bf k})\Pi
 _{nn^{\prime }}^{\beta }({\bf k})}{%
\omega -\omega _{nn^{\prime }}({\bf k})+i\gamma }\,,  
\label{defsig}
\end{eqnarray}
where $f(\epsilon _{n{\bf k}})$ is the Fermi function; $\hbar \omega
_{nn^{\prime }}({\bf k})\equiv \epsilon _{n{\bf k}}-\epsilon _{n^{\prime }
  {\bf k}}$ is the energy difference of Kohn-Sham energies; $\gamma $ is the
lifetime parameter, describing the finite lifetime of the excited Bloch
electron states; $\Pi _{nn^{\prime }}^{\alpha }$ are the dipole optical
transition matrix elements. \cite{book:AHY04} A detailed description of the
optical matrix elements is given in Refs. \onlinecite{ABP+93,book:AHY04}. The
absorptive part of the optical conductivity was calculated in a wide energy
range. The Kramers-Kronig transformation was then used to calculate the
dispersive parts of the optical conductivity from the absorptive part.  We
used the value $\gamma$=0.6 eV for the interband relaxation parameter.

Within the one-particle approximation, the absorption coefficient
$\mu^{\lambda}_j (\omega)$ for incident x-ray of polarization $\lambda$ and
photon energy $\hbar \omega$ can be determined as the probability of
electronic transitions from initial core states with the total angular
momentum $j$ to final unoccupied Bloch states

\begin{eqnarray}
\mu^j_{\lambda} (\omega) &=& \sum_{m_j} \sum_{n \bf k} | \langle \Psi_{n \bf k} |
\Pi _{\lambda} | \Psi_{jm_j} \rangle |^2 \delta (E _{n \bf k} - E_{jm_j} -
\hbar \omega ) \nonumber \\
&&\times \theta (E _{n \bf k} - E_{F} ) \, ,
\label{mu}
\end{eqnarray}
where $\Psi _{jm_j}$ and $E _{jm_j}$ are the wave function and the energy of a
core state with the projection of the total angular momentum $m_j$;
$\Psi_{n\bf k}$ and $E _{n \bf k}$ are the wave function and the energy of a
valence state in the $n$-th band with the wave vector {\bf k}; $\ef$ is the
Fermi energy.

$\Pi _{\lambda}$ is the electron-photon interaction operator in the dipole
approximation

\begin{equation}
\Pi _{\lambda} = -e \mbox{\boldmath$\alpha $} \bf {a_{\lambda}},
\label{Pi}
\end{equation}
where $\bm{\alpha}$ are the Dirac matrices, $\bf {a_{\lambda}}$ is the
$\lambda$ polarization unit vector of the photon vector potential, with
$a_{\pm} = 1/\sqrt{2} (1, \pm i, 0), a_{\parallel}=(0,0,1)$. Here, $+$ and $-$
denotes, respectively, the left and right circular photon polarizations with
respect to the magnetization direction in the solid. Then, x-ray magnetic
circular and linear dichroism are given by $\mu_{+}-\mu_{-}$ and
$\mu_{\parallel}-(\mu_{+}+\mu_{-})/2$, respectively.  More detailed
expressions of the matrix elements for the spin-polarized fully relativistic
LMTO method may be found in Refs.~\onlinecite{GET+94,ABP+93}.

The Eliashberg function (the spectral function of the electron-phonon
interaction) expressed in terms of the phonon linewidths
$\gamma_{\mathbf{q}\nu}$ has the form \cite{Allen72}

\begin{equation}
\alpha^2F(\omega) = \frac{1}{2\pi N(\epsilon_F)}\sum_{\mathbf{q}\nu}
\frac{\gamma_{\mathbf{q}\nu}}{\omega_{\mathbf{q}\nu}}\delta(\omega
-\omega_{\mathbf{q}\nu}) ,
\label{mu_Bgr}
\end{equation}

The line-widths characterize the partial contribution of each phonon:

\begin{equation}
\gamma_{\mathbf{q}\nu}= 2\pi\omega_{\mathbf{q}\nu}\sum_{jj'
  \mathbf{k}} | g_{ \mathbf{k}+\mathbf{q}j',
  \mathbf{k}j}^{\mathbf{q}\nu} |^2 \delta (\epsilon _{j \mathbf{k}} -
\epsilon_{F}) \delta (\epsilon _{\mathbf{k}+\mathbf{q}j'} -
\epsilon_{F}).
\label{nu}
\end{equation}

The electron-phonon interaction constant is defined as:

\begin{equation}
\lambda_{e-ph} = 2\int_0^\infty\frac{d\omega}{\omega}{\alpha^2}F(\omega) ,
\label{lamda}
\end{equation}

It can also be expressed in terms of the phonons line-widths:

\begin{equation}
\lambda_{e-ph} = \sum_{ \mathbf{q}\nu}\frac{\gamma_{ {\bf q}\nu}}{\pi N
  (\epsilon_F) \omega_{ {\bf q}\nu}^2},
\label{lamda2}
\end{equation}
were N($\epsilon_F$) is the electron density of states per atom and per spin
on the Fermi level ($\epsilon_F$) and $g_{ \mathbf{k}+ {\bf q}j'
  \mathbf{k}j}^{ {\bf q}\nu}$ is the electron-phonon interaction matrix
element.  The double summation over Fermi surface in Eq.(\ref{nu}) was carried
out on dense mesh (793 point in the irreducible part of the BZ)

Calculations of the electronic structure and physical properties of the
TiB$_2$ and ZrB$_2$ diborides were performed using a scalar relativistic
FP-LMTO method \cite{Sav96} with the experimentally observed lattice
constants: $a$=3.167 \AA\, and $c$=3.529 \AA\, for ZrB$_2$; \cite{StPe75}
$a$=3.03 \AA\, $c$=3.229 \AA, for TiB$_2$. \cite{OtIs94} For the LMTO
calculations we used the Perdew-Wang \cite{PW92} parameterization of the
exchange-correlation potential in general gradient approximation.  BZ
integrations were performed using the improved tetrahedron
method. \cite{BJA94} Phonon spectra and electron-phonon matrix elements were
calculated for 50 points in the irreducible part of the BZ using the linear
response scheme developed by Savrasov.  \cite{Sav96} The 3$s$ and 3$p$
semi-core states of TiB$_2$ were treated as valence states in separate energy
windows (for ZrB$_2$ : 4$s$ and 4$p$). Variations in charge density and
potential were expanded in spherical harmonics inside the MT sphere as well as
2894 plane waves in the interstitial area with 88.57 Ry cut-off energy for
ZrB$_2$ and 97.94 Ry cut-off energy for TiB$_2$. As for the area inside the MT
spheres, we used 3k$-spd$ LMTO basis set energy (-0.1, -1, -2.5 Ry) with
one-center expansions inside the MT-spheres performed up to $l_{max}$ = 6.

\section{\label{sec:results}Results and discussion}

\subsection{Energy band structure}

\begin{figure}[tbp!]
\begin{center}
\includegraphics[width=0.45\textwidth]{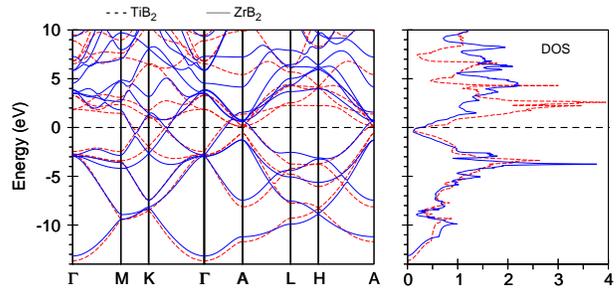}
\end{center}
\caption{\label{Ek} (Color online) Energy band structure and total
  DOS [in states/(cell eV)] of ZrB$_2$ (full blue lines) and TiB$_2$
  (dashed red lines).}
\end{figure}

\begin{figure}[tbp!]
\begin{center}
\includegraphics[width=0.45\textwidth]{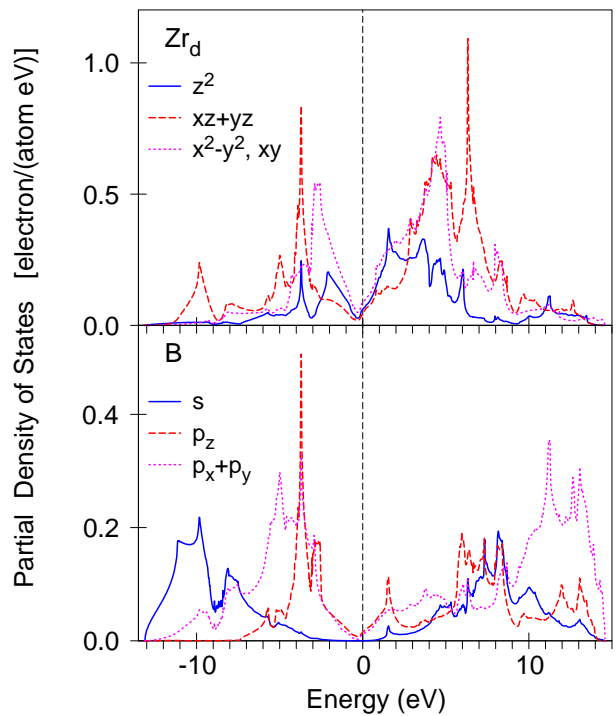}
\end{center}
\caption{\label{PDOS} (Color online) Partial DOSs [in states/(atom
    eV)] of ZrB$_2$.}
\end{figure}

Figure~\ref{Ek} presents the energy band structure and total density of states
(DOS) of ZrB$_2$ (full lines) and TiB$_2$ (dashed lines). The partial DOSs
ZrB$_2$ are shown in Fig. \ref{PDOS}. Our results for the electronic structure
of ZrB$_2$ and TiB$_2$ are in agreement with earlier calculations.
\cite{Singh04,GFL+09,FGP+09,DCC09,ZLL+09,DCC10,FTH+10} A common feature for
all transition metal diborides is the deep DOS minimum (pseudo-gap) at the
Fermi energy separating the valence band and the conduction band. According to
Pasturel {\it et al.}, \cite{PCH85} a pseudo-gap arises because of a strong
chemical interaction. The M-B covalent bonding is believed to be responsible
for this effect. Fig. \ref{Ek} includes a comparison of the total DOS for
ZrB$_2$ and TiB$_2$. In both systems, we observe a deep minimum in the DOS at
the Fermi energy, although the gap appears slightly broader in the case of
ZrB$_2$. The Zr 4$d$ states in ZrB$_2$ are the dominant features in the
interval from $-$12.5 eV to 9 eV. These tightly bound states show overlap with
B 2$p$ and, to a lesser extent, with B 2$s$ states both above and below $\ef$,
implying considerable covalency. Higher-energy states between 9 eV and 20 eV
above $\ef$ appear to arise from Zr 5$p$ and 6$s$ states hybridized with B
2$p$ states. The crystal field at the Zr site ($D6h$ point symmetry) causes
the splitting of Zr $d$ orbitals into a singlet $a_{1g}$ ($\dzz$) and two
doublets $e_{1g}$ ($\dyz$ and $\dxz$) and $e_{2g}$ ($\dxy$ and $\dxxyy$). The
crystal field at the B site ($D3h$ point symmetry) causes the splitting of B
$p$ orbitals into a singlet $a_4$ ($p_z$) and a doublet $e_2$ ($p_x$ and
$p_y$). B $s$ states occupy a bottom of valence band between $-$13.1 eV and
$-$3.0 eV and hybridize strongly with B $p_x$ and $p_y$ and Zr $\dyz$ and
$\dxz$ states located at $-$12.5 eV to $-$0.5 eV. B $p_x$ and $p_y$ states are
located between $-$12.5 eV and $-$0.5 eV.  B $p_z$ states occupied a smaller
energy interval from $-$7.5 eV to $-$0.5 eV with a very strong and narrow peak
structure at around $-$4 eV.

\subsection{X-ray absorption and photoemission spectra}

Experimentally the electronic structure of ZrB$_2$ and TiB$_2$ has been
investigated by means of photoemission spectroscopy, \cite{IHN77,TEK+03} point
contact spectroscopy, \cite{NKY+02} x-ray absorption spectroscopy,
\cite{TEK+03,CWA+05} and optical spectroscopy. \cite{Linton74,RDA+08,RAC+08}

\begin{figure}[tbp!]
\begin{center}
\includegraphics[width=0.45\textwidth]{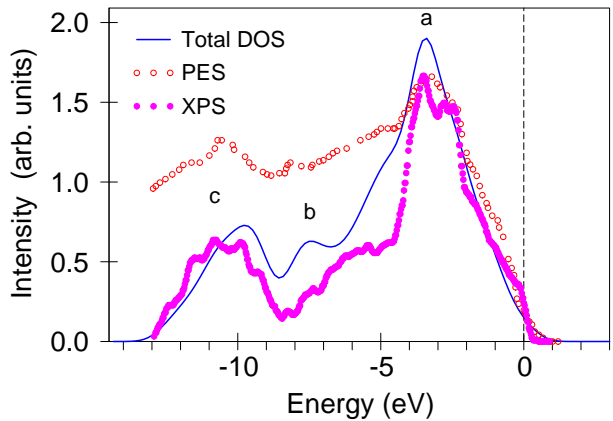}
\end{center}
\caption{\label{XPS} (Color online) Comparison of the total DOS (full
  line) with photoemission \protect\cite{TEK+03} (open circles) and
  x-ray photoemission \protect\cite{IHN77} (full circles) of ZrB$_2$.}
\end{figure}

Figure \ref{XPS} shows the experimentally measured photoemission (PES)
\cite{TEK+03} and x-ray photoemission (XPS) \cite{IHN77} spectra of ZrB$_2$
compared with the calculated energy distribution of total DOS. The calculated
DOS has been broadened to account for life-time effects and for the
experimental resolution. The characteristic features of the XPS are divided
into three parts ranging from the Fermi energy to $-$4.5 eV (peak $a$), $-$4.5
eV to $-$8.5 eV (peak $b$), and $-$8.5 eV to $-$13 eV (peak $c$). The low
energy peak $c$ arises mostly from the B 2$s$ states and partly from the low
energy peak of Zr 4$d_{xz,yz}$ states (see Fig. \ref{PDOS}). The major peak
$a$ close to the Fermi energy is derived by Zr 4$d$ states. B $p$ states as
well as the Zr 4$d_{xz,yz}$ states contribute to the broad peak $c$ located
from $-$4.5 eV to $-$8.5 eV. Agreement between experiment and theory in energy
position of major fine structures is reasonably well. However, peak $b$ is
slightly shifted toward lower energy in the theory, besides, peak $a$ does not
split into two peaks as observed in the experimental XPS spectrum. On the
other hand, the experimental PE spectrum \cite{TEK+03} measured at 325.26 eV
(open circles in Fig. \ref{XPS}) has a single peak $a$ in close agreement with
the theoretically calculated DOS. Intensity of the low energy part of the PE
spectrum is significantly increased due to inelastically scattered
electrons. The corresponding background was extracted from the experimental
XPS spectrum. \cite{IHN77} It is interesting to note that the position of the
peak $b$ in DOS is in better agreement with the PE spectrum than observed in
the case of the XPS spectrum.

\begin{figure}[tbp!]
\begin{center}
\includegraphics[width=0.45\textwidth]{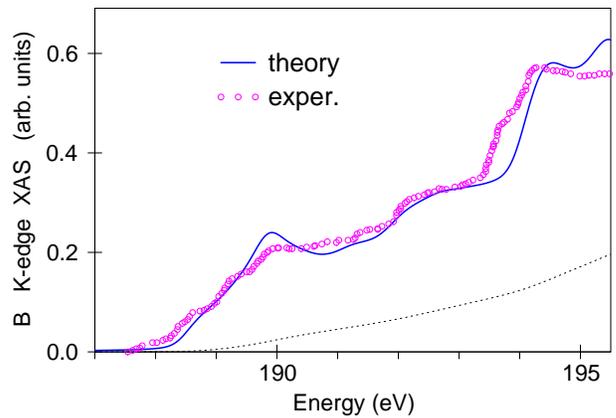}
\end{center}
\caption{\label{B_K} (Color online) The theoretically calculated and
  experimentally measured \protect\cite{TEK+03} x-ray absorption spectra at
  B $K$ edge of ZrB$_2$.}
\end{figure}

\begin{figure}[tbp!]
\begin{center}
\includegraphics[width=0.45\textwidth]{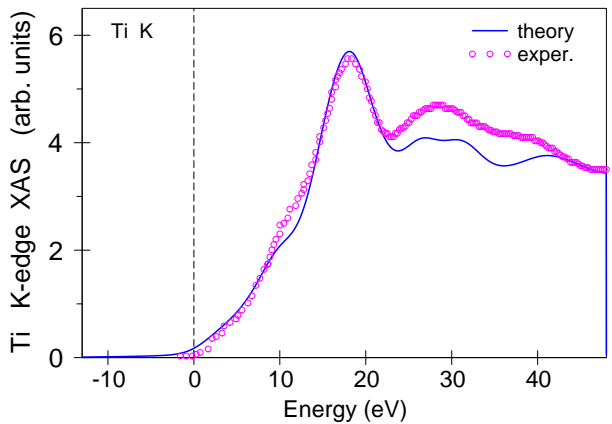}
\end{center}
\caption{\label{Ti_K} (Color online) The theoretically calculated and
  experimentally measured \protect\cite{CWA+05} x-ray absorption
  spectra at Ti $K$ edge of TiB$_2$.}
\end{figure}

X-ray absorption spectra (XAS) were measured by Tsuda {\it et al.}
\cite{TEK+03} at the B $K$ and Zr $M_{2,3}$ edges of ZrB$_2$. Ti $K$ XAS were
measured by Chu {\it et al.}  \cite{CWA+05} The XA spectra in metals at the
$K$ edge in which the 1$s$ core electrons are excited to the $p$ states
through the dipolar transition usually attract only minor interest because $p$
states are not the states of influencing magnetic or orbital order. Recently,
however, understanding $p$ states has become important due to XA spectroscopy
using $K$ edges of transition metals gaining popularity. The $K$ edge XAS is
sensitive to electronic structures at neighboring sites because of the
delocalized nature of the $p$ states.

\begin{figure}[tbp!]
\begin{center}
\includegraphics[width=0.45\textwidth]{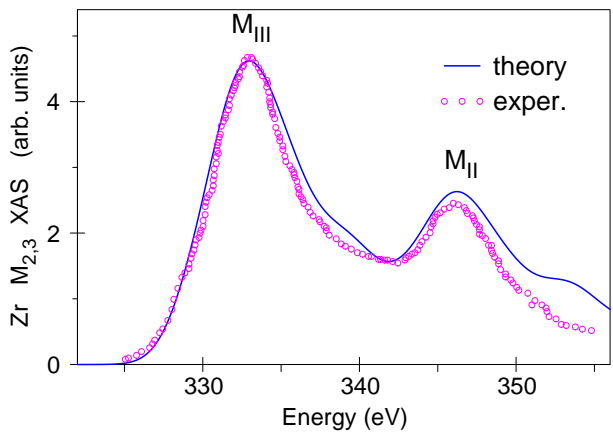}
\end{center}
\caption{\label{Zr_M23} (Color online) The theoretically calculated and
  experimentally measured \protect\cite{TEK+03} x-ray absorption spectra at
  Zr $M_{2,3}$ edges of ZrB$_2$. }
\end{figure}

Figure \ref{B_K} presents the theoretically calculated and experimentally
measured B $K$ XPS spectra ($1s\to2p$ transitions). The agreement between the
theory and the experiment is excellent. The low energy peak around 190 eV is
due to transitions from the 1$s$ core level to the mostly B $p_z$ states (see
Fig. \ref{PDOS}) with some amount of the $p_x$ and $p_y$ states. Fine
structure at 194.5 eV reflects the corresponding peak between 6 eV and 7 eV
above the Fermi level (Fig. \ref{PDOS}). Figure \ref{Ti_K} presents
theoretically calculated and experimentally measured \cite{CWA+05} Ti $K$ XPS
spectra in TiB$_2$. The agreement between theory and experiment is also quite
good; except for a second major peak around 28 eV that is slightly
underestimated theoretically.

Figure \ref{Zr_M23} presents theoretically calculated and experimentally
measured Zr $M_{2,3}$ XPS spectra ($3p\to4d$ transitions). Agreement between
theory and experiment is again good. Because of the dipole selection rules
(apart from the 4$s_{1/2}$-states that which have a small contribution to the
XAS due to relatively small 3$p$ $\to$ 5$s$ matrix elements \cite{book:AHY04})
only 3$d_{3/2}$-states occur as final states for $M_2$ XAS. For the $M_3$ XAS,
4$d_{5/2}$-states also contribute. Although the 3$p_{3/2}$ $\to$ 4$d_{3/2}$
radial matrix elements are only slightly smaller than for the 3$p_{3/2}$ $\to$
4$d_{5/2}$ transitions, the angular matrix elements strongly suppress the
3$p_{3/2}$ $\to$ 4$d_{3/2}$ contribution. \cite{book:AHY04} Therefore in
neglecting the energy dependence of the radial matrix elements, the $M_2$ and
the $M_3$ spectra can be viewed as a direct mapping of the DOS curve for
4$d_{3/2}$- and 4$d_{5/2}$-character, respectively.

\subsection{Optical spectra}

\begin{figure}[tbp!]
\begin{center}
\includegraphics[width=0.5\textwidth]{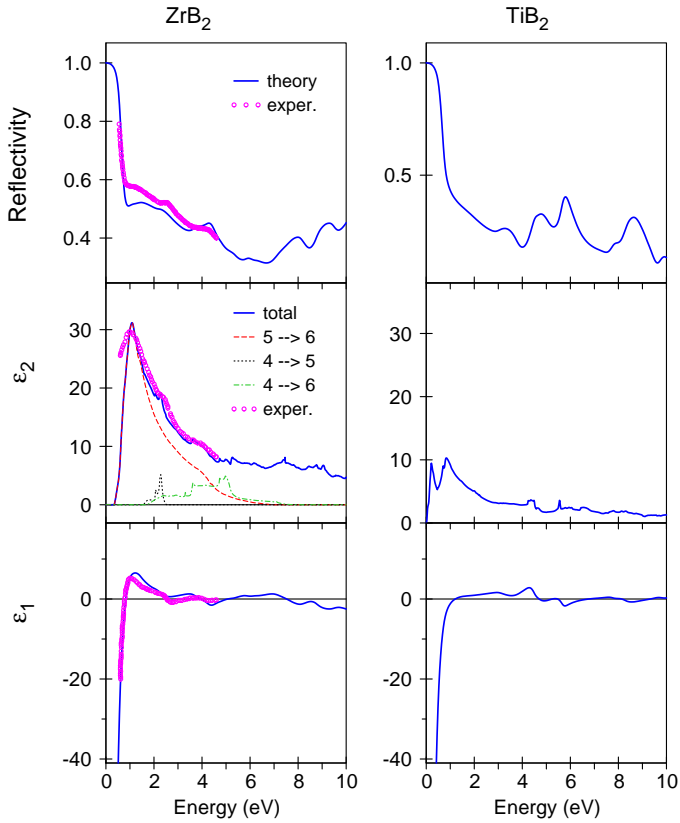}
\end{center}
\caption{\label{optics} (Color online) Theoretically calculated (solid blue
  lines) and experimentally measured (open circles) \protect\cite{RAC+08}
  optical reflectivity spectra (upper panel) and dielectric constants
  $\varepsilon_2$ (middle panel) and $\varepsilon_1$ (lower panel) of
  ZrB$_2$. For the $\varepsilon_1$ function the contributions of different
  interband transitions are presented. }
\end{figure}

The optical spectra of ZrB$_2$ have been measured by several
authors. \cite{Linton74,RDA+08,RAC+08} Currently there are no such
measurements for TiB$_2$. Fig. \ref{optics} shows the theoretically calculated
and experimentally measured optical reflectivity spectra $R(\omega)$ as well
as dielectric constants $\varepsilon_1(\omega)$ and $\varepsilon_2(\omega)$
for ZrB$_2$. Also presented are theoretically calculated $R(\omega)$,
$\varepsilon_1(\omega)$, and $\varepsilon_2(\omega)$ for TiB$_2$. Theory
reproduces well peculiarities of ZrB$_2$ optical spectra.

\begin{figure}[tbp!]
\begin{center}
\includegraphics[width=0.45\textwidth]{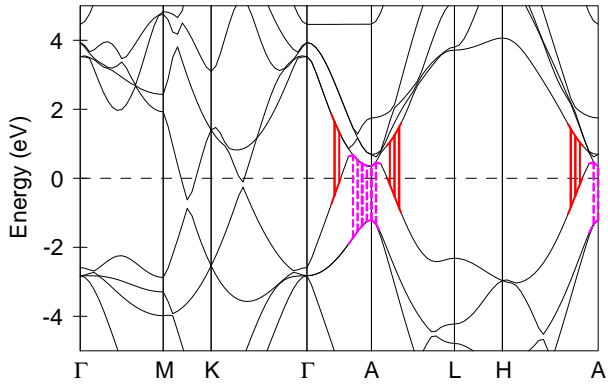}
\end{center}
\caption{\label{O_interband} (Color online) Theoretically calculated 4 $\to$ 5
  (pink dotted lines) and 5 $\to$ 6 (red full lines) interband transitions in
  the optical conductivity of of ZrB$_2$. }
\end{figure}

We performed decomposition of the calculated $\varepsilon_2$ spectrum into the
contributions arising from separate interband transitions and different places
of {\bf k} space. We found that the major peak in the $\varepsilon_2(\omega)$
(around 1 eV) is mostly determined by the 5 $\to$ 6 interband transitions
along the $\Gamma-$A and A$-$L symmetry directions
(Fig. \ref{O_interband}). The shoulder at 2 eV is due to the 4 $\to$ 5
interband transitions around A symmetry point (pink dotted lines in
Fig. \ref{O_interband}).

Although the band structures ZrB$_2$ and TiB$_2$ are very similar (see
Fig. \ref{Ek}), their optical spectra visibly differ from each other. The
experimental measurements of the optical spectra of TiB$_2$ are highly
desirable.

\subsection{Fermi surface}

The magnetoresistance and Hall effect were measured in early investigations of
the Fermi surface (FS) of ZrB$_2$ in 1966 by Piper. \cite{Pip66} He showed
that ZrB$_2$ is a compensated semimetal with an effective concentration of
0.04 electrons/cell with no open trajectories. In 1978, the dHvA effect was
observed and investigated by Tanaka {\it et al.}  \cite{TIB+78} and an attempt
was made to interpret the dHvA oscillations on the basis of calculations of
the band structure of CrB$_2$ \cite{LKE+75} using the "rigid band"
approximation. The model obtained for the FS of ZrB$_2$ was later confirmed by
improved calculations performed using the FLAPW method. \cite{RAP+02} Recently
the Fermi surfaces of ScB$_2$, ZrB$_2$ and HfB$_2$, were studied by Pluzhnikov
{\it et al.} \cite{PSD+07} using the dHvA effect. Their results for ZrB$_2$
are similar to previous measurements by Tanaka. \cite{TIB+78}

\begin{figure}[tbp!]
\begin{center}
\includegraphics[width=0.45\textwidth]{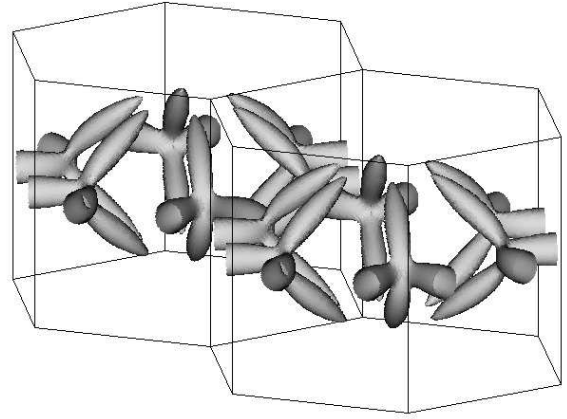}
\end{center}
\caption{\label{FS_el} (Color online) The calculated electron sheets of
  the Fermi surface around K symmetry point from the 6th energy band
  of ZrB$_2$.}
\end{figure}

\begin{figure}[tbp!]
\begin{center}
\includegraphics[width=0.45\textwidth]{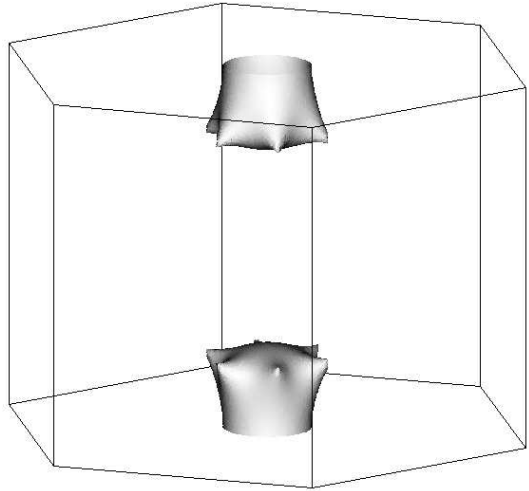}
\end{center}
\caption{\label{FS_hole} The calculated hole sheets of the Fermi surface
  at the A symmetry point from the 5th energy band of ZrB$_2$.}
\end{figure}

Theoretical calculations show a ring-like electron FS around the $K$ symmetry
point (Fig. \ref{FS_el}) and of a wrinkled dumbell-like hole FS at the A point
(Fig. \ref{FS_hole}) in ZrB$_2$.  The electron FS and hole FS have threefold
and sixfold symmetries, respectively.  These are broadly consistent with the
Fermi surfaces used by Tanaka \cite{TIB+78} to interpret their dHvA
data. TiB$_2$ has very similar sheets of its Fermi surface.

\begin{figure}[tbp!]
\includegraphics[width=0.3\textwidth]{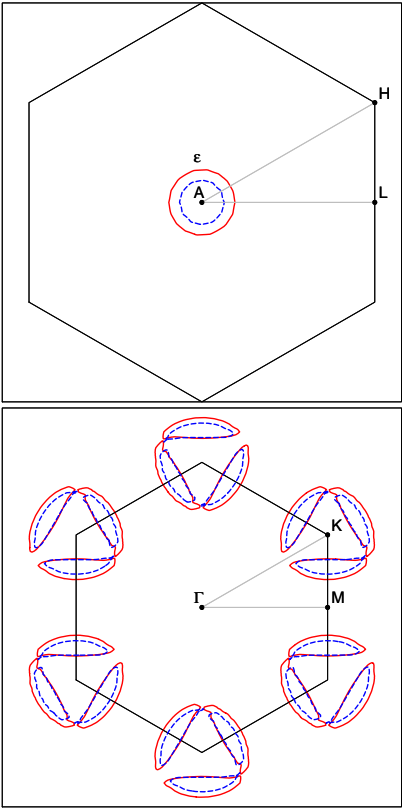}
\caption{\label{FS_cs} (Color online) The calculated cross sections in
  the plane perpendicular $z$ direction and crossed $A$ symmetry point
  (upper panel) and $\Gamma$ point (lower panel) for ZrB$_2$ (full red
  curves) and TiB$_2$ (dashed blue curves). The labels are provided as
  used in the text. }
\end{figure}

Figure \ref{FS_cs} shows the calculated cross section areas in the plane
perpendicular $z$ direction and crossed $A$ symmetry point for hole FS (upper
panel) and $\Gamma$ point for electron FS (lower panel) of ZrB$_2$ and
TiB$_2$. It can be clearly seen that TiB$_2$ has a smaller FS than ZrB$_2$.

\begin{figure}[tbp!]
\begin{center}
\includegraphics[width=0.45\textwidth]{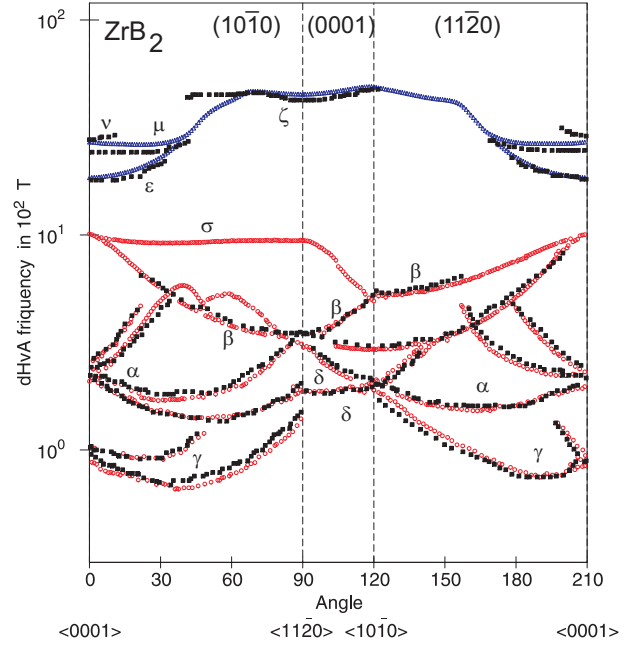}
\end{center}
\caption{\label{FS_Zr_S} (Color online) The calculated (open red and blue
  circles for the electron and hole surfaces, respectively) and experimentally
  measured \protect\cite{PSD+07} (black full squares) angular dependence of
  the dHvA oscillation frequencies in the compound ZrB$_2$.}
\end{figure}

Figure \ref{FS_Zr_S} represents angular variations of the experimentally
measured dHvA frequencies \cite{PSD+07} for ZrB$_2$ in comparison with the
first-principle calculations for field direction in the ($10\bar10$),
($11\bar20$), and (0001) planes. The observed frequencies of $\alpha$,
$\beta$, $\gamma$, and $\delta$ oscillations belong to electron FS around the
$K$ point (see Fig. 4 in Ref. \onlinecite{TIB+78}). The $\epsilon$, $\nu$,
$\mu$, and $\zeta$ orbits belong to the hole wrinkled dumbbell FS.  The
$\alpha$ frequencies have four branches at the ($10\bar10$) plane and three
branches at the ($11\bar20$) plane. The lower $\gamma$ frequencies have two
branches in both the planes.

The theory reasonably well reproduces the frequencies measured experimentally.
However, there are still some discrepancies. For high frequencies in the
$<0001>$ direction, we found the $\epsilon$ and $\mu$ branches but were unable
to obtain $\nu$ branch. We also discover a new branch $\sigma$ which is not
detected experimentally. This branch belongs to the electron FS around the $K$
point. It has almost a constant frequency at the $(10\bar10)$ plane and
rapidly drops in frequency at the (0001) plane. The theoretically calculated
$\zeta$ orbits exist in a wider angle interval than observed experimentally.

\begin{figure}[tbp!]
\begin{center}
\includegraphics[width=0.45\textwidth]{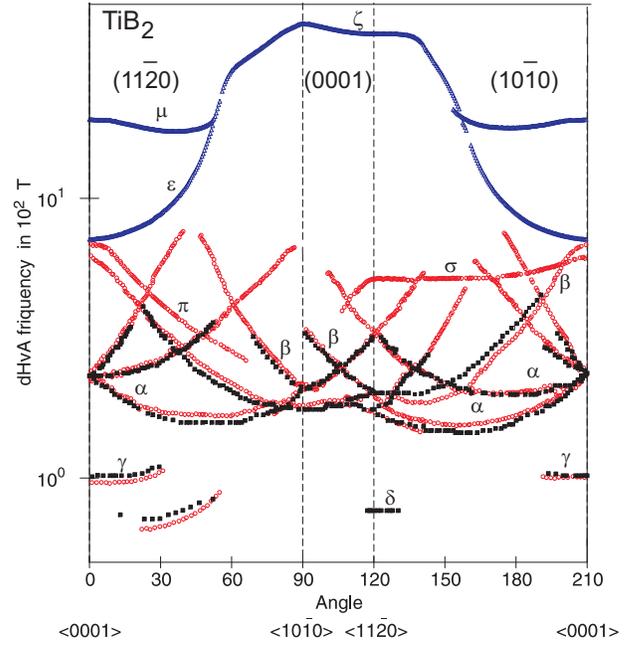}
\end{center}
\caption{\label{FS_Ti_S} (Color online) The calculated (open red and blue
  circles for the electron and hole surfaces, respectively)
  and experimentally measured \protect\cite{TaIs+80} (black full
  squares) angular dependence of the dHvA oscillation frequencies in
  the compound TiB$_2$.}
\end{figure}

\begin{figure}[tbp!]
\begin{center}
\includegraphics[width=0.45\textwidth]{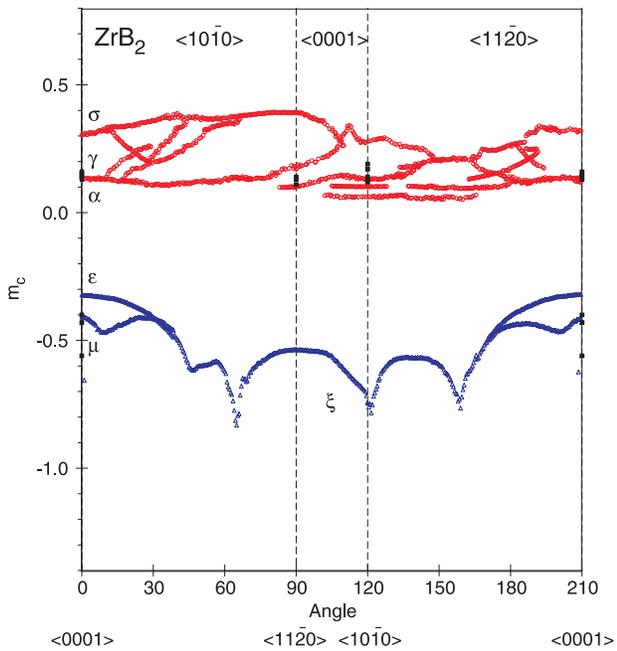}
\end{center}
\caption{\label{FS_Zr_mc} (Color online) The calculated angular dependence of
  the cyclotron masses for the electron Fermi surface (open red circles) and
  the hole Fermi surface (blue open triangles) and experimentally measured
  ones (black full squares) in the compound ZrB$_2$. }
\end{figure}

\begin{figure}[tbp!]
\begin{center}
\includegraphics[width=0.45\textwidth]{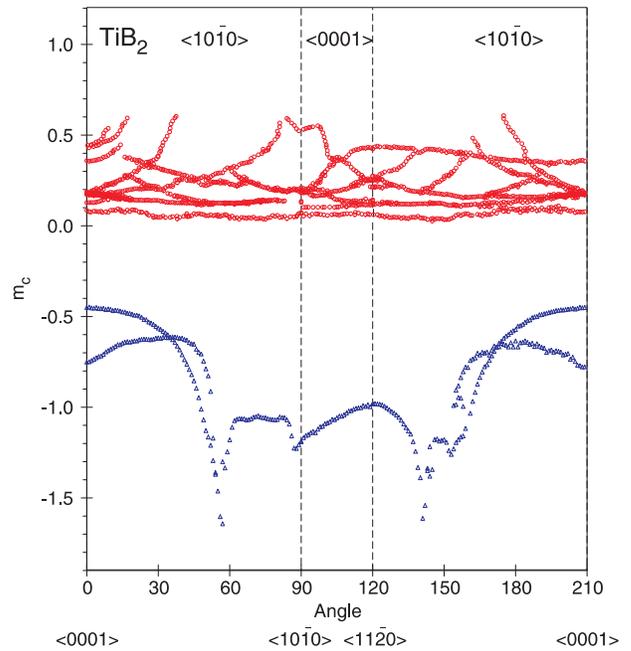}
\end{center}
\caption{\label{FS_Ti_mc} (Color online) The calculated angular dependence of
  the cyclotron masses for the electron Fermi surface (red open circles) and
  the hole Fermi surface (blue open triangles) in the compound TiB$_2$. }
\end{figure}

Figure \ref{FS_Ti_S} represents an angular variation of experimentally
measured dHvA frequencies \cite{TaIs+80} in TiB$_2$ when compared with
theoretically calculated frequencies. The theoretical calculations quite well
reproduce the angle dependence of the extremal cross sections for low
frequency orbits $\gamma$, $\alpha$ and $\beta$. Similar to ZrB$_2$ we
detected theoretically a new branch $\sigma$ in TiB$_2$ which is not observed
experimentally. This branch belongs to the electron FS around the $K$
point. We also find an additional orbit $\pi$ at the ($11\bar20$) plane which
is absent in ZrB$_2$ and did not detected experimentally. We were not able to
find theoretically low frequency $\delta$ oscillations appeared in a small
angle interval near the $<11\bar20>$ direction. For high frequencies we found
the $\epsilon$, $\mu$ and $\zeta$ branches similar to the corresponding orbits
in ZrB$_2$. However, these orbits have not been detected in the dHvA
experiment. \cite{TaIs+80} One of the possible reasons for that is the
relatively large cyclotron masses for these orbits. Figures \ref{FS_Zr_mc} and
\ref{FS_Ti_mc} show the calculated angular dependence of the cyclotron masses
for ZrB$_2$ and TiB$_2$, respectively. The cyclotron masses for the
$\epsilon$, $\mu$, and $\zeta$ orbits in TiB$_2$ are much higher than the
corresponding orbits in ZrB$_2$. The masses for the low-frequency oscillations
$\alpha$, $\beta$, $\gamma$ and $\delta$ are less than 0.2$m_0$ for ZrB$_2$
and slightly larger in TiB$_2$.

\subsection{Phonon spectra}

\begin{figure}[tbp!]
\begin{center}
\includegraphics[width=0.45\textwidth]{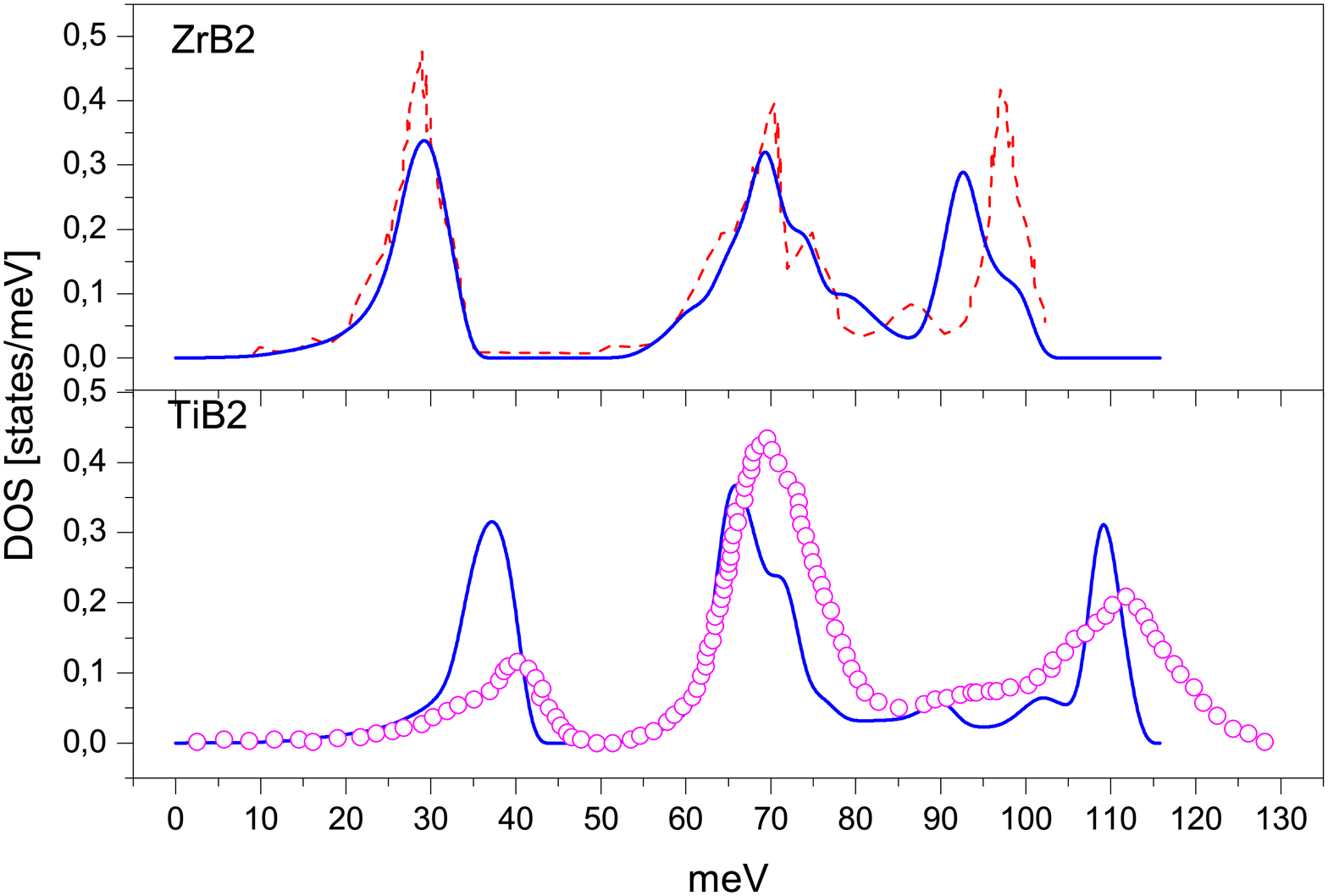}
\end{center}
\caption{\label{P_DOS} (Color online) Theoretically calculated phonon
  density of states (full blue lines) for ZrB$_2$ and TiB$_2$ and
  experimentally measured one for TiB$_2$ \protect\cite{HRS+03} (open
  circles). Dashed red line presents the calculated phonon DOS of
  ZrB$_2$ by Deligoz {\it et al.} [\protect\onlinecite{DCC10}]. }
\end{figure}

The unit cell of TB$_2$ (T=Zr, Ti) contains three atoms, which give in general
case a nine phonon branches.  Figure \ref{P_DOS} shows theoretically
calculated phonon density of state for ZrB$_2$ and TiB$_2$. The DOS for both
ZrB$_2$ and TiB$_2$ can be separated into three distinct regions.  Based on
our analysis of relative directions of eigenvectors for each atom in unit
cell, we find that the first region (with a peak in phonon DOS at 29 meV in
ZrB$_2$ and 37,5 meV in TiB$_2$) is dominated by the motion of the
transition-metal atoms Zr and Ti, respectively. This region belongs to the
acoustic phonon modes. The shift of the first region in the phonon DOS towards
lower frequencies for ZrB$_2$ in comparison to TiB$_2$ is due to the higher
mass of Zr.  The second wide region (60-80 meV) results from the coupled
motion of Zr(Ti) and the two B atoms in the unit cell. The $E_{1u}$, $A_{2g}$,
$B_{1g}$ phonon modes (see Table \ref{fre}) lie in this area. The phonon DOS
in the third region extends from 88 meV to 103 meV in ZrB$_2$ and from 105 meV
to 115 meV in TiB$_2$. This is due to the movement of boron atoms and is
expected since boron is lighter than transition metal atoms. The covalent
character of the B-B bonding is also crucial for the high frequency of
phonons.  The in-plane E$_{2g}$ mode belongs to this region. The second and
third regions represent optical phonon modes in crystals. The most significant
feature in the phonon DOS is a gap around 40 to 60 meV for both ZrB$_2$ and
TiB$_2$. This gap is a consequence of the large mass difference, which leads
to decoupling of transition metal and boron vibrations.

\begin{table}[tbp!]
  \caption{Theoretically calculated phonon frequencies (in meV) in the
    $\Gamma$ symmetry point for ZrB$_2$ and TiB$_2$ and experimentally
    measured ones for TiB$_2$ \protect\cite{HRS+03} as well as calculated
    phonon frequencies in ZrB$_2$ calculated by Deligoz {\it et al.}
    Ref. [\protect\onlinecite{DCC10}].}
 
\label{fre}
\begin{tabular}{|c|c|c|c|c|c|}
\hline
Compound	&    reference      & $E_{1u}$ & $A_{2g}$ & $B_{1g}$ & $E_{2g}$ \\
\hline
& our results  & 58.70 & 63.26 & 71.0 & 99.70 \\
ZrB$_2$ & Ref. [\onlinecite{DCC10}] & 60.61 & 63.49 & 67.76 &  98.45  \\ 
		        
\hline

& our results  &63 .0& 63.5 & 69.1 & 110.0 \\
TiB$_2$  & Ref. [\onlinecite{HRS+03}] & 65.5  & 66.4 & 70.0 & 112.8 \\
		
\hline

\end{tabular}
\end{table}

The TiB$_2$ phonon DOS was measured using inelastic neutron scattering
experiments in Ref. \onlinecite{HRS+03}. Our results are in good agreement
with the experiment (see Fig. \ref{P_DOS}, lower panel). The small discrepancy
in the positions of main peaks for TiB$_2$ does not exceed accuracy of
calculation.

Currently, there are no data concerning the experimentally measured phonon DOS
in ZrB$_2$. So we compare our results with theoretically calculated phonon DOS
by Deligoz {\it et al.} \cite{DCC10} (Fig. \ref{P_DOS}, upper
panel). Calculations of these authors were based on the density functional
formalism and generalized gradient approximation. They used the
Perdew-Burke-Ernzerhof functional \cite{Perd96} for the exchange-correlation
energy as it is implemented in the SIESTA code. \cite{Ord96,Sol02} This code
calculates the total energies and atomic Hellmann-Feynman forces using a
linear combination of atomic orbitals as the basis set. The basis set consists
of finite range pseudoatomic orbitals of the Sankey-Niklewsky type
\cite{San89} generalized to include multiplezeta decays.  The interactions
between electrons and core ions are simulated with the separable
Troullier-Martins \cite{Tro91} normconserving pseudopotentials. In other
words, they used the so-called "frozen phonon" technique and built an
optimized rhombohedral supercell with 36 atoms. This method is inconvenient
for calculating phonon spectra for small {\bf q}-points as well as for
compounds with large number of atoms per unit cell. There is a very good
agreement between our calculations and the results of Deligoz {\it et al.}
\cite{DCC10} in a shape and energy position of two first low energy peaks in
the phonon DOS. There is only a low energy shift of the third peak by $\sim$5
meV in our calculations in comparison with results of Deligoz {\it et al.}
\cite{DCC10} (see also Table I).

\subsection{Electron-phonon interaction}

\begin{figure}[tbp!]
\begin{center}
\includegraphics[width=0.45\textwidth]{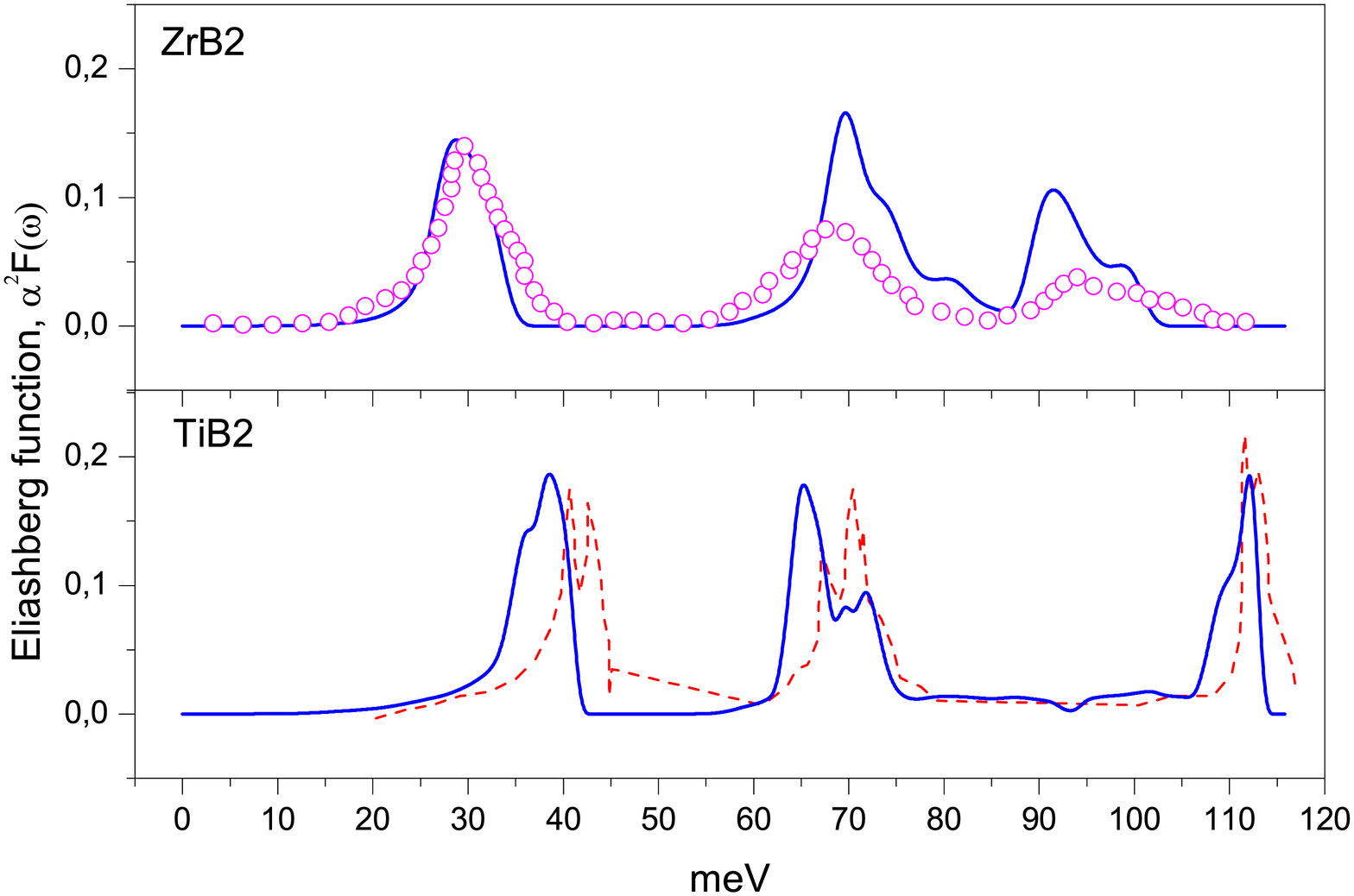}
\end{center}
\caption{\label{Eli} (Color online) The theoretically calculated Eliashberg
  function $\alpha F(\omega)$ of ZrB$_2$ and Ti2B$_2$ (full blue lines) and
  experimentally measured point contact spectral function
  \protect\cite{NKY+02} (open circles) for ZrB$_2$. Dashed red line presents
  Eliashberg function of TiB$_2$ calculated by Heid {\it et al.}
  Ref. [\protect\onlinecite{HRS+03}]. }
\end{figure}

Figure \ref{Eli} shows theoretically calculated Eliashberg functions for
ZrB$_2$ and TiB$_2$. We find no significant difference in the shape and energy
position of major peaks between phonon DOS values and electron-phonon coupling
functions in these compounds. Therefore, we can conclude that electron-phonon
Eliashberg function is mostly defined by the shape of phonon DOSs in ZrB$_2$
and TiB$_2$. There are no regions with unusually high electron-phonon
interaction and phonon dispersion curves do not contain any soft modes which
might be indicative of the possible superconductivity in these borides. By
integrating the Eliashberg function using equation (\ref{lamda}), we estimate
the average electron-phonon interaction constant to be $\lambda_{e-ph}$=0.14
for ZrB$_2$. A similar result was obtained earlier by Singh \cite{Singh04}
($\lambda$=0.15). Drechsler {\it et al.} \cite {DRO+04} estimated the value of
the dHvA orbit averaged el-ph coupling constant to be $\lambda \le$0.1. A weak
electron-phonon coupling strength of $\lambda \sim$0.1 was derived from both
the comparison of the calculated density of states at the Fermi level and
specific heat data (Fuchs {\it et al.}  \cite{FDM+03}), and by point-contact
measurements ($\lambda_{PC}$=0.06 \cite{NKY+02}).

Figure \ref{Eli} (upper panel) represents the PC electron-phonon interaction
function for ZrB$_2$ in comparison with the theoretically calculated Eliashber
function. Results closely agree in the energy positions of major
peaks. However, the experimental PC function displays a monotonically
decreasing peak amplitude (as we move along the energy scale in the
high-energy direction). As a consequence, the coupling PC constant
$\lambda_{PC}$=0.06 is less than that obtained from the integration of the
Eliashberg function ($\lambda_{e-ph}$=0.14). The disagreement might be
explained by the fact that PC and the Eliashberg functions have a slightly
different nature. First, the kinematic restriction of electron scattering
processes in a PC is taken into account by a factor $K = \frac{1}{2} (1 -
\theta \tan \theta)$, where $\theta$ is the angle between initial and final
momenta of scattered electrons (for the Eliashberg function, the corresponding
factor $K$=1). Therefore in PC spectra the large angle ($\theta \to \pi$)
backscattering processes are dominated. The second reason for suppressing
high-energy peaks in the PC function is a deviation from the ballistic
electron flow in point-contact spectroscopy. (PC spectra can not be described
in the framework of ballistic regime for a high-energy phonon area).
\cite{NKY+02}

For TiB$_2$ we again obtain a small value of the electron-phonon constant
$\lambda_{e-ph}$=0.15. Due to absence of the experimentally measured
electron-phonon spectral function in TiB$_2$ we compare our calculations with
theoretical results obtained by Heid \cite{HRS+03} who used the mixed basis
pseudopotential method. \cite{Boh01,He02} There is relatively good agreement
between our calculations and Heid's results for the energy position and shape
of the peaks (Fig. \ref{Eli}). The first two major low energy peaks of the
Eliashberg function are slightly shifted towards the smaller energies in
comparison with the results of Heid. \cite{HRS+03}

\subsection{Electrical resistivity}

In the pure metals (excluding low-temperature region), the electron-phonon
interaction is the dominant factor governing electrical conductivity of the
substance. Using lowest-order variational approximation, the solution for the
Boltzmann equation gives the following formula for the temperature dependence
of $\rho_{I}(T)$:

\begin{equation}
\rho_I (T) = \frac{\pi\Omega_{cell} k_B T}{N(\epsilon_F)\langle
  v_I^2\rangle}\int_0^\infty\frac{d\omega}{\omega}\frac{\xi^2}{sinh^2\xi}{\alpha_{tr}^2}F(\omega),
\label{resist}
\end{equation}
where, the subscript $I$ specifies the direction of the electrical current. In
our work, we investigate two direction: [0001] (c-axis or z direction) and
[10$\bar1$0] (a-axis or x-direction).  $\langle v_I^2\rangle$ is the average
square of the $I$ component of the Fermi velocity, $\xi=\omega/{2k_BT}$.

Mathematically, the transport function $\alpha_{tr} F(\omega)$ differs from
$\alpha F(\omega)$ only by an additional factor
$[1-v_{I}(\mathbf{k})v_{I}(\mathbf{k}')/\langle v_I^2\rangle]$, which
preferentially weights the backscattering processes.

 Formula (\ref{resist}) remains valid in the range $\Theta_{tr}/5 < T <
 2\Theta_{tr}$ \cite{Sav96} where:

\begin{equation}
\Theta_{tr} \equiv \langle\omega^2\rangle_{tr}^{1/2} ,
\label{debay}
\end{equation}

\begin{equation}
\langle\omega^2\rangle_{tr}=
\frac{2}{\lambda_{tr}}\int_0^\infty\omega\alpha_{tr}^2
F(\omega)d\omega ,
\label{deb2}
\end{equation}

\begin{equation}
\lambda_{tr}=2 \int_0^\infty\alpha_{tr}^2 F(\omega)\frac{d\omega}{\omega} ,
\label{deb3}
\end{equation}

The low-temperature electrical resistivity is the result of electron-electron
interaction, size effects, scattering on impurities, etc., however, for high
temperatures it is necessarily to take into account the effects of anharmonicity
and the temperature smearing of the Fermi surface. The $\Theta_{tr}$=604.8 K
and 646.19 K for ZrB$_2$ and TiB$_2$, respectively.

\begin{figure}[tbp!]
\begin{center}
\includegraphics[width=0.45\textwidth]{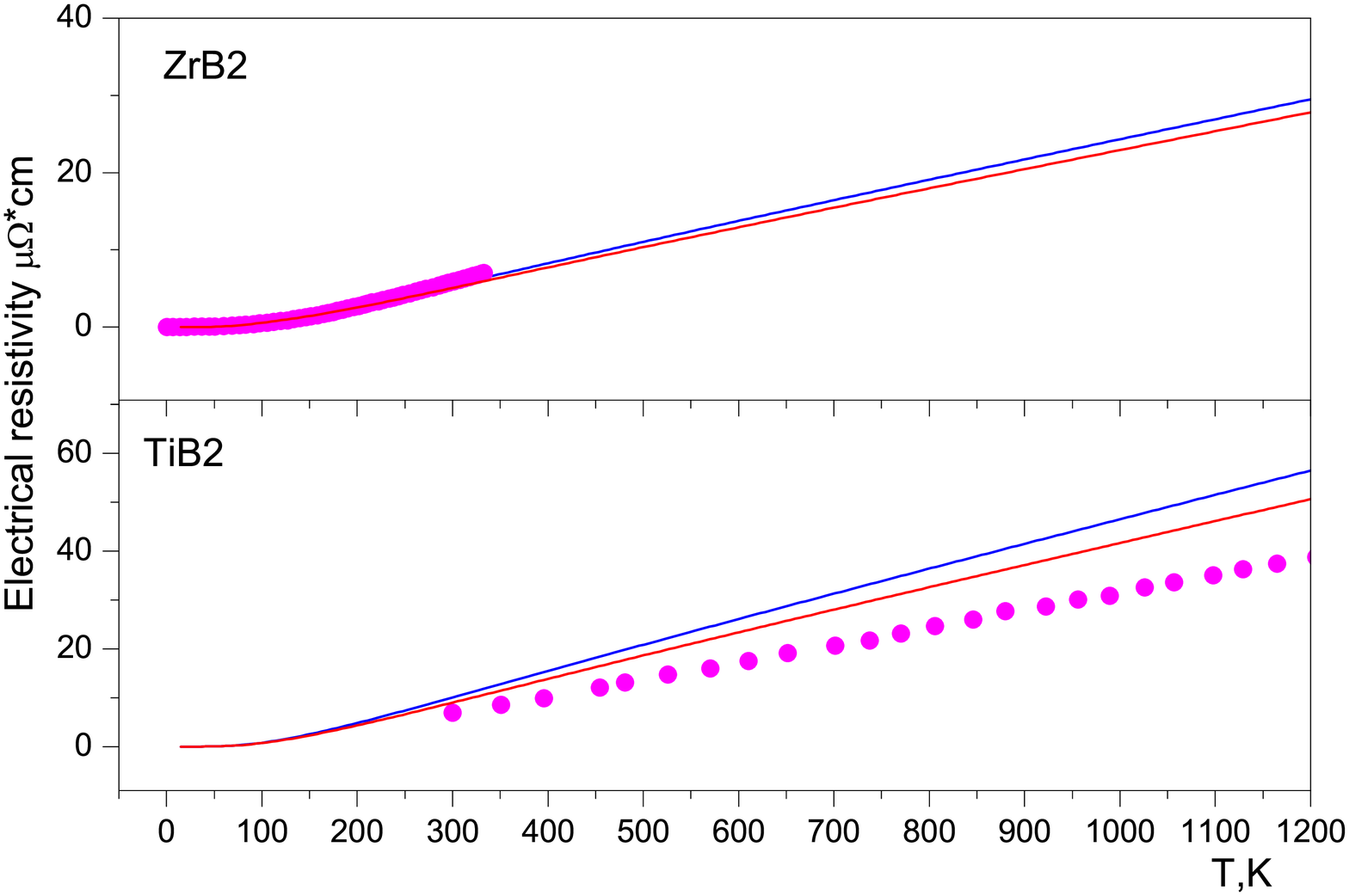}
\end{center}
\caption{\label{P3} (Color online) Theoretically calculated for the $<0001>$
  direction (blue curves) and the basal $<10\bar10>$ direction (red curves)
  and experimentally measured temperature dependence of electrical resistivity
  of ZrB$_2$ \protect\cite{Mcl84} (upper panel) and TiB$_2$
  \protect\cite{Mcl84} (lower panel). }
\end{figure}

Figure \ref{P3} represents the experimental data for mono-crystalline ZrB$_2$
\cite{Mcl84} as well as our calculations (upper panel). No evidence of
anisotropy of the electrical resistivity was found experimentally. Our
theoretical calculations also show quite small anisotropically behavior of the
electrical resistivity in ZrB$_2$ (compare red and blue curves in
Fig. \ref{P3}, upper panel). There is a good agreement between our
calculations and experimentally measured results in the region up to 350 K.

We found that the anisotropy of the electrical resistivity in TiB$_2$
(Fig. \ref{P3}, lower panel) is larger than it was in ZrB$_2$. Our theoretical
results slightly exceed experimental data, \cite{Mcl84} especially at high
temperatures.  This is due to using in our calculations the lowest-order
variational approximation in solution of the Boltzmann equation which gives
upper limit for the electrical resistivity.  \cite{book:Zim60,Allen72}

\section{\label{sec:summ}Summary}

We have studied the electronic structure and physical properties of ZrB$_2$
and TiB$_2$ using a full potential linear muffin-tin orbital method. We
investigated the electron and phonon subsystems as well as the electron-phonon
interaction in these compounds. The theory shows good agreement with
experimentally measured x-ray absorption spectra at the B and Ti $K$ and Zr
$M_{2,3}$ edges. Agreement between the experiment and the theory in optical
spectra of ZrB$_2$ is also good. We found that the major peak in the
$\varepsilon_2(\omega)$ of ZrB$_2$ around 1 eV is mostly determined by the 5
$\to$ 6 interband transitions along $\Gamma-$A and A$-$L symmetry directions.

We investigated the Fermi surface, angle dependence of the cyclotron masses,
and extremal cross sections of the Fermi surface of ZrB$_2$ and TiB$_2$ in
details. Theoretical calculations show a ring-like electron FS in ZrB$_2$
around the $K$ symmetry point and a wrinkled dumbbell-like hole FS at the A
point. TiB$_2$ has a smaller FS than ZrB$_2$. Theory reproduces the
experimentally measured dHvA frequencies in both the ZrB$_2$ and TiB$_2$
reasonably well. We found that masses for low-frequency oscillations $\alpha$,
$\beta$, $\gamma$, and $\delta$ are less than 0.2$m_0$. Masses for
high-frequency oscillations $\epsilon$, $\nu$, $\mu$, and $\zeta$ are large.
We discover new branches $\sigma$ both in ZrB$_2$ and TiB$_2$ which did not
detected experimentally.  Theoretical calculations closely reproduce the angle
dependence of the extremal cross sections of high frequency orbits $\epsilon$,
$\mu$, and $\zeta$ in ZrB$_2$. Similar orbits appeared in the theoretical
results for TiB$_2$, but not detected experimentally. The cyclotron masses for
these orbits in TiB$_2$ are much higher than the corresponding orbits in
ZrB$_2$ (compare Figs. (\ref{FS_Zr_mc}) and (\ref{FS_Ti_mc})). It could be one
of the reasons why they have not been observed in the dHvA measurements.
\cite{TaIs+80}

Calculated phonon spectra and phonon DOSs for both ZrB$_2$ and TiB$_2$ are in
good agreement with experimental results as well as previous calculations. The
Elishberg function of electron-phonon interaction in ZrB$_2$ is in good
agreement with the experimentally measured point contact spectral function for
both the position and the shape of the major peaks. We did not find regions
with high electron-phonon interaction or phonon dispersion curves with soft
modes in either ZrB$_2$ or TiB$_2$. This is in agreement with the fact that no
trace of superconductivity was found in these borides. The averaged
electron-phonon interaction constant was found to be rather small
$\lambda_{e-ph}$=0.14 and 0.15 for ZrB$_2$ and TiB$_2$, respectively. We
calculated the temperature dependence of the electrical resistivity in ZrB$_2$
and TiB$_2$ in the lowest-order variational approximation of the Boltzmann
equation. We found rather small anisotropical behavior of the electrical
resistivity in ZrB$_2$ to be in good agreement with experimental
observation. We found that the anisotropy of electrical resistivity in TiB$_2$
is larger than it is in ZrB$_2$.

\section*{Acknowledgments}

This work was carried out at the Ames Laboratory, which is operated for the
U.S.Department of Energy by Iowa State University under Contract
No. DE-AC02-07CH11358. This work was supported by the Director for Energy
Research, Office of Basic Energy Sciences of the U.S. Department of
Energy. This work was also supported by the National Academy of Sciences of
Ukraine in the framework of the State Target Scientific and Technology Program
of Implementation and Application of Grid Technologies for
2009-2013. V.N.A. gratefully acknowledges the hospitality during his stay at
Ames Laboratory.


\newcommand{\noopsort}[1]{} \newcommand{\printfirst}[2]{#1}
  \newcommand{\singleletter}[1]{#1} \newcommand{\switchargs}[2]{#2#1}

\end{document}